\documentclass[12pt,dvips]{article}

\usepackage{amsmath,amssymb,exscale}
\usepackage{array,multicol}
\usepackage{afterpage,float,flafter}
\usepackage{epsfig,rotating,pifont}
\usepackage{cite}
\@addtoreset{equation}{section} \makeatother
\renewcommand{\theequation}{\thesection.\arabic{equation}}
\def\beq{\begin{eqnarray}}
\def\eeq{\end{eqnarray}}

\def\lsim{\mathrel{\rlap{\lower3pt\hbox{\hskip0pt$\sim$}}
   \raise1pt\hbox{$<$}}}         
\def\gsim{\mathrel{\rlap{\lower4pt\hbox{\hskip1pt$\sim$}}
   \raise1pt\hbox{$>$}}}         

\title{
\vspace{0cm}
\huge{Gaugino Mass in AdS space}
\vspace*{0.7cm}
\author{
\Large {\text{Ben Gripaios$^{*\dagger}$, Hyung Do Kim$^{\dagger \ddagger}$, Riccardo Rattazzi$^\dagger$,}}\\
\Large{\text{Michele Redi$^\dagger$ and
Claudio A. Scrucca$^\dagger$}}\\ \\\emph{$^*$CERN PH-TH, CH-1211 Geneva 23, Switzerland}\\ 
\emph{$^\dagger$ITPP, EPFL, CH-1015, Lausanne, Switzerland}\\
\emph{$^\ddagger$ FPRD and Department of Physics and Astronomy, SNU, Seoul, Korea, 151-747}
}
}

\date{}
\begin{document}
\maketitle \thispagestyle{empty} \vspace*{-.2cm}

\begin{abstract}
We study supersymmetric QED in AdS$_4$ with massless matter. At 1-loop the ultra-violet
regulator of the theory generates a contribution to the gaugino
mass that is na\"{\i}vely inconsistent with unbroken
supersymmetry. We show that this effect, known in flat space as
anomaly mediated supersymmetry breaking, is required to cancel 
an  infra-red contribution
arising from the boundary conditions in AdS space, which necessarily 
break chiral symmetry. We also discuss an analogous  UV/IR cancellation that is  independent of supersymmetry.

\end{abstract}

\newpage
\renewcommand{\thepage}{\arabic{page}}
\setcounter{page}{1}

\section{Introduction}
\label{introduction}

In phenomenologically interesting models the effects of broken
supersymmetry in the visible sector are conveniently
parameterized, working in an off-shell formulation, by the
expectation values of the auxiliary components of some hidden
sector supermultiplets. Among the auxiliary fields,  the scalar
$u$, belonging to the graviton supermultiplet, $(g_{\mu\nu},
\psi_\mu^\alpha, A_\mu, u)$, stands out as  special. Indeed,
unlike for auxiliary fields belonging to matter and gauge hidden
sector  multiplets, the coupling of $u$ is completely fixed (at
the leading relevant order) once the masses and self-couplings of
the low energy effective theory, prior to supersymmetry breaking,
are specified. This property just follows from $u$ being a partner
of  $g_{\mu\nu}$ whose coupling is equally well specified by the
energy momentum tensor of the low energy effective theory. The
scenario of `Anomaly Mediated' (AM) supersymmetry breaking
corresponds to the limiting case in which the contribution of $u$
dominates over all of the others \cite{am1,am2}. The name `Anomaly
Mediated' is due to the fact that in the MSSM $u$ only couples to
the visible fields at the quantum level, via a supersymmetric
analogue of the dilatation anomaly of non-supersymmetric field
theory.

The purpose of this paper will not be to build phenomenological models based
on AM, but rather to investigate some of its more amusing theoretical
aspects. In fact, far away from the domain of phenomenology, 
we shall be working in four dimensional supersymmetric Anti-de-Sitter (AdS) space. 
We nonetheless believe that our study provides interesting additional insight 
into the properties of AM, in particular its being UV insensitive, in spite 
of being UV generated.

To set the stage, it is convenient to derive AM terms via the superconformal approach
to supergravity \cite{ferrara}. At tree level, the most general two-derivative 
Lagrangian may be written as
\begin{equation}
{\cal L}= \Bigl [S^{\dagger} S
\,\Omega(\Phi^{\dagger}_i,e^{q_i V} \Phi_i)\Bigr ]_D+  \left \{\Bigl [S^3
W(\Phi_i)+ f(\Phi_i) W^{\alpha}W_{\alpha})\Bigr ]_F +{{\rm h.c.}}\right \} ,
\end{equation}
where $D$ and $F$ are superconformally invariant densities, provided
that the chiral superfield, $S$, and the matter fields, $\Phi_i$, have
Weyl  weights 1 and 0, respectively.  Interesting actions are
obtained by consistently taking the lowest component of $S$ with
non-vanishing expectation value. This breaks the superconformal
group down to Poincar\'e supergravity and turns $S$  into a purely
auxiliary field, formally restoring scale invariance, hence the
name `superconformal compensator'. Indeed a suitable
superconformal gauge can be chosen where $S=1+\theta^2 u$.  The
couplings of the auxiliary field $u$ are thus fixed by dilations
and R-symmetry. In particular a classically scale invariant
subsector, like the MSSM, couples to $u$ only at the quantum
level. For a massless gauge theory the coupling of $S$ is easily
read off by demanding formal scale- (and R-) invariance of the 1PI
action at 1-loop
\begin{equation}
{\Gamma} = \frac 1 4 \left [W^{\alpha}\Bigl( \frac{1}{g^2(\mu)}+\frac {b}{8\pi^2} \ln
(\frac {\sqrt{\Box}}{\mu S})\Bigr )  W_{\alpha}\right ]_F +h.c.
\end{equation}
By expanding in components, one finds a gaugino mass term which 
is proportional to the $\beta$-function
\begin{equation}
m_\lambda=-\frac {b g^2}{16\pi^2} u.
\label{gauginomass}
\end{equation}
The dependence of $\Gamma$ on $S$ is local, compatibly with its
being UV generated. However, it belongs to a non-local supergravity
invariant `structure' (involving $\ln \Box$), and this is why it
is convenient to use the 1PI action to determine it. This is just
the supersymmetric generalization of a dilaton coupling to the
trace anomaly, hence the name `anomaly mediation'.

In models with broken supersymmetry and vanishing cosmological constant,  
$\langle u \rangle =O(m_{3/2})$, implying a 1-loop contribution of order 
$ (\alpha/4\pi)m_{3/2}$ to gaugino and sfermion masses. However, 
one may also have $\langle u \rangle \not =0$, with unbroken supersymmetry
on AdS.  In that case, the expectation  value is given by the superpotential: 
$\langle u\rangle =W/M_P^2=1/L$, where $L$ is the AdS radius.
Indeed, at tree level, $\langle u\rangle = 1/L$ generates the mass
splittings, of order of the AdS curvature, that are required by supersymmetry
in AdS. The r\^{o}le of a loop effect like anomaly mediation is
less clear in this case, though it ought to be easy to understand,
given that the theory still enjoys unbroken supersymmetry.

The purpose of this note is to explain the r\^{o}le played by
anomaly mediation in supersymmetric AdS. This issue was 
briefly considered in \cite{ds}, in the context of a general discussion 
in which the short distance origin of AM was emphasized. 
However our explanation for the r\^{o}le of AM in AdS space differs 
from the one proposed in \cite{ds}.
We will argue that the existence of AM is  a necessary consequence
of supersymmetry, given the large-distance properties of AdS space, 
in particular the presence of a (conformal) boundary.
In this sense, our work represents yet another way of deriving AM
masses, purely via consideration of IR saturated quantities.
The outline is as follows. In section \ref{adssusy}, we review supersymmetry 
in AdS and supersymmetric QED therein. In section \ref{mass}, we compute
the 1-loop contributions to the gaugino self-energy in SQED with massless 
matter, and discuss the implications for the gaugino mass. In section \ref{out}, 
we present conclusions. The case of SQED with massive matter is relegated 
to the appendix.

\section{Supersymmetry in AdS Space}
\label{adssusy}

In this section, we briefly review some basic features of
supersymmetry in four-dimensional AdS space which will be relevant
for the following discussion. For more details, see \cite{breit,Nicolai:1984hb} 
and refs.\ therein.

The isometry group of AdS$_4$ is $SO(2,3)$, whose unitary,
infinite-dimensional representations are denoted by $D(E,s)$,
where $E$ and $s$ represent respectively the energy and spin of
the lowest energy state in the representation. The Lagrangian mass
parameter of the corresponding fields (in units of $1/L$) are
functions of $E$ and $s$. For instance, for the simplest cases of
$s=0,\frac{1}{2}$, we have
\begin{eqnarray}
D(E,0)\qquad\longrightarrow \qquad m_0^2&=&\frac {E(E-3)}{L^2},\label{scalar} \\
D(E,\frac{1}{2})\qquad\longrightarrow \qquad m_{\frac 1 2}^2&=& \frac {(E-{3}/{2})^2}{L^2}\label{fermion}\, .
\end{eqnarray}
Just as in flat space, the simplest irreducible representations of
the super-group $Osp(1,4)$ correspond to  chiral and vector
supermultiplets. A chiral supermultiplet decomposes into the following
representations of $SO(2,3)$:
\begin{equation}
D\left(E_0,0\right)\oplus D\left(E_0+\frac 1 2,\frac 1 2\right)\oplus
D\left(E_0+1,0\right),~~~~~~~~~~E_0\ge \frac 1 2 \label{chiral}.
\end{equation}
Note that the supersymmetry generators raise and lower $E$ 
by a half-integer.  Then, according to eqs.~(\ref{scalar}),(\ref{fermion}),  
the mass terms for fermions and scalars  within the same supermultiplet 
are not, in general, the same. These splittings are mandated by $Osp(1,4)$ and
originate within the lagrangian from two sources. One source is the non-vanishing 
Ricci scalar and the other source is $\langle u\rangle = 1/L$. Notice, finally, 
that in the special case of the conformally-coupled supermultiplet, with $E_0=1$, 
the two scalars have the same mass, even though they belong to different 
representations: namely $D(1,0)$ and $D(2,0)$.

Turning now to the massless vector supermultiplet,  the
$SO(2,3)$ representation content is
\begin{equation}
D\left(\frac 3 2 ,\frac 1 2\right)\oplus D\left(2 , 1\right).
\label{vector}
\end{equation}
This multiplet is both conformally coupled and `short',
corresponding to its being related to a gauge invariant
lagrangian. A massive vector multiplet, on the other hand, is
characterized by $E_0>3/2$, and decomposes as
\begin{equation}
D\left( E_0 ,\frac 1 2\right)\oplus D\left(E_0+
\frac 1 2 , 0\right)\oplus D\left(E_0+\frac 1 2 , 1\right)\oplus \left (E_0+1,\frac 1 2 \right)\, .
\end{equation}
This is a long multiplet that can be viewed as arising from a
Higgs mechanism. Indeed, it has the same state multiplicity as the
direct sum of the massless vector supermultiplet and the Goldstone
supermultiplet, whose content is $D\left(2 , 0\right)\oplus D\left(\frac 5 2 ,
\frac 1 2 \right)\oplus \left (3,0 \right)$. Since it corresponds
to multiplet shortening, the masslessness condition must be stable in
perturbation theory. In particular, the gaugino mass, for an
unbroken gauge symmetry, must be zero to all orders.

\subsection{AdS SUSY QED}

The presence of the anomaly mediated contribution to the mass 
(\ref{gauginomass}) is, na\"{\i}vely, at odds with the previous observation that the gaugino should be massless. 
To clarify the r\^{o}le of AM, we shall focus on the simplest non-trivial example, 
that is  the mass of the gaugino in supersymmetric QED.
Our theory consists of $\mathcal{N}=1$ supergravity with a vector superfield $V$, and
two chiral superfields $\Phi_{\pm}$, with opposite charges $\pm1$. The K\"ahler and superpotential functions are given by
(throughout the paper we use the conventions of Wess and Bagger  \cite{ws})
\begin{eqnarray}
\Omega\equiv -3M_P^2e^{-K/3M_P^2}&=&-3M_P^2+ \Phi^{\dagger}_+ e^{gV} \Phi_+ +\Phi^{\dagger}_- e^{-gV} \Phi_-+O(\Phi^4), \nonumber \\
W&=&\frac {M_p^2}{L}+ m \Phi_+ \Phi_-,\\
f&=& 1+O(\Phi_+\Phi_-).
\end{eqnarray}
Since we shall be working in the neighbourhood of $\Phi_\pm=0$, we neglect the higher order
terms indicated by $O(\dots)$. The constant term in the superpotential 
gives rise to the AdS$_4$ background and to the expectation value of the compensator,
\begin{equation}
\langle S\rangle = 1+\frac{1}{L}\theta^2\,.
\label{compensator}
\end{equation}
We will find it technically convenient to work  in the Poincar\'{e} patch, with metric
\begin{equation}
ds^2=\frac {L^2}{z^2} \left(dx^{\mu}dx_{\mu}+dz^2\right).
\label{AdS$_4$}
\end{equation}
The co-ordinates $x^{\mu}$  ($\mu=0,1,2$) and $z$ cover only one of an infinite set of similar 
Poincar\'e patches of the full AdS space. However Poincar\'e co-ordinates cover the whole euclidean 
AdS (EAdS), which can be obtained just by the substitution $t\to i\tau$ (see for instance the discussion 
in ref. \cite{Aharony:1999ti}). This last property indicates that, if properly interpreted, computations
on the Poincar\'e patch yield informations about the properties of QFT on full AdS.
Assuming $L$ to be positive, in these co-ordinates the four unbroken supersymmetries 
are parameterized by the Killing spinors
\begin{equation}
\xi= z^{\frac1 2}[\epsilon_0-i \sigma^3 \bar{\epsilon}_0]+z^{-1/2} x_\mu \sigma^\mu [\epsilon_0+i \sigma^3 \bar{\epsilon}_0],
\label{unbrokensusy}
\end{equation}
where $\epsilon_0$ is a two-component constant spinor. Notice that the Killing spinors
naturally decompose into two real spinors of $SO(1,2)$. The first of these corresponds to the standard $\mathcal{N}=1$ 
in 2+1 dimensions, while the other corresponds to the conformal supersymmetry. In fact, for our purposes it will
suffice to consider the flat supersymmetries, as the others are implied by the AdS isometries.

By taking the limit $M_P\to\infty$ with $L$ fixed, we decouple gravity 
and focus on quantum effects that are purely due to SQED on AdS$_4$. The relevant Lagrangian is, therefore,
\begin{eqnarray}
{\cal L/}\sqrt g&=& \left [{\rm kinetic}+{\rm gauge\, D\, terms}\right  ] -m (\psi_+\psi_-+\bar{\psi}_+\bar{\psi}_-)\nonumber \\
&-&(m^2 - \frac 2 {L^2}) (|\phi_+|^2+|\phi_-|^2)+\frac  m  L (\phi_+\phi_- + \phi_+^*\phi_-^*) \nonumber \\
&+& i g \sqrt{2}\lambda(\psi_+ \phi_+^*-\psi_-\phi_-^*)- i g \sqrt{2}\bar{\lambda}(\bar{\psi}_+ \phi_+-\bar{\psi}_- \phi_-),
\end{eqnarray}
where, without loss of generality, we have taken $m$ to be real.
One sees that the scalars acquire non-holomorphic mass terms,
originating from the non-vanishing Ricci scalar, and holomorphic
(B-type) masses, arising from the compensator F-term. (The
fermionic mass and interaction terms, by contrast, retain the same form as  in flat
space.) The scalar mass eigenstates and their masses are given by
\begin{eqnarray}
\phi_{1,2} & = & \frac 1 {\sqrt{2}}(\phi_+\mp \phi_-^*),\\
m_{1,2}^2 & = & \frac{1}{L^2} \left( -2\pm m L + (mL)^2\right).
\label{combination}
\end{eqnarray}

Eqs.~(\ref{gauginomass}),(\ref{compensator}) imply the presence of 
an AM contribution to the gaugino mass, given by
\begin{equation}
\Delta_{UV} {\cal L}= - \frac {g^2}{16\pi^2 L}\lambda \lambda+ h.c.\equiv -\frac{1}{2}m_{UV} \lambda\lambda + h.c.\,.
\label{gauginobilinear}
\end{equation}
As explained above and emphasized in \cite{ds}, a gaugino mass would be incompatible 
with supersymmetry in AdS$_4$. Indeed, for $m\not = 0$, there is 
an additional contribution to $m_\lambda$, corresponding to a finite threshold 
effect at the scale $m$, where matter is integrated out. This is due to the presence of both a
fermion mass and an R-breaking B-type mass for the scalars. By the
well known property of AM in flat space, we can directly conclude
that, at least for  $mL\gg 1$, the threshold effect cancels
eq.~(\ref{gauginobilinear}), at least up to subleading effects of $O(1/mL)$.
However, it would be nice to see the exact cancellation in an
explicit computation. Moreover, in the limit $m=0$, corresponding to conformal multiplets,
there seems to be a puzzle, in that all sources of R-symmetry breaking disappear from
the matter lagrangian! In other words, for $m=0$
there is, at first sight, no obvious contribution in addition to eq. (\ref{gauginobilinear}).
In \cite{ds}, it was concluded that the contribution in eq. (\ref{gauginobilinear})
does  not affect  the physical mass (defined in the sense of the
representation of AdS), since $g^2$   runs to zero in the
infrared. This explanation is, however, puzzling, as it
requires an all-orders resummation of diagrams, while we expect the
supersymmetry algebra to be satisfied at each finite order in
perturbation theory. Furthermore, this argument cannot be applied
to the non-Abelian case. In actual fact, the resolution of the gaugino 
mass puzzle has to do with the boundary conditions in AdS, which 
shall be discussed in the next section. What we shall find there is that 
boundary effects provide a calculable, IR saturated, contribution to the 
gaugino bilinear in the 1-loop 1PI effective action. This contribution corresponds to 
a mass $m_{IR}$ which exactly  cancels the UV one
\begin{equation}
m_{UV}+m_{IR}=0\, .
\label{uvir}
\end{equation}

\subsection{Boundary conditions}

The most relevant feature of AdS space, for our discussion, is the
presence of a (conformal) boundary located at $z=0$ in the Poincar\'e patch (\ref{AdS$_4$}). 
One immediate consequence of the presence of a 2+1-dimensional boundary  is that
chiral symmetry is always broken in AdS$_4$ \cite{allen}. This is fully analogous to what happens in a field theory on
half of flat space: when a fermion travelling towards the boundary is reflected, the momentum flips 
sign, while $J_z$ is conserved. Thus, helicity is not conserved. 

More formally, chiral symmetry is broken by the  boundary conditions that are necessary 
to define the theory. This can be seen by considering a two component spinor propagating on half of flat space, with action
\begin{equation}
S=\frac 1 2\int_{z\geq 0} d^4 x \left[ \bigl (-i \psi \sigma^m D_m \bar{\psi}-
m  \psi\psi\bigr ) +{\rm {h.c.}}\right].
\label{flat}
\end{equation}
The variation of the action is
\begin{equation}
\delta S= (EOM)-\frac i 2 \left [ \delta \psi \sigma^3 \bar{\psi}-{\rm {h.c.}}\right]_{z=0}.
\end{equation}
In order to obtain sensible boundary conditions ({\em i.e.} not over-constraining), 
a boundary term
$-\frac{1}{4}\int_{z=0}e^{-i\varphi} \ \psi \psi +{\rm {h.c.}}$
must be added to the action, where $\varphi$ is an arbitrary phase.
The variational principle then demands that
\begin{equation}
\psi_{\alpha}\Big\vert_{z=0}=
ie^{i\varphi} \sigma^3_{\alpha\dot{\alpha}}\bar{\psi}^{\dot{\alpha}}\Big\vert_{z=0},
\label{flatfermion}
\end{equation}
implying that chiral symmetry is broken even for vanishing bulk
mass\footnote{For $m=0$, without loss of generality one can choose $\varphi=0$.}.

The generalization to AdS requires some care, because of the 
divergent scale factor at $z=0$. The boundary conditions in this case can 
be derived by considering the behavior of the solutions close to $z=0$.
Without loss of generality, we can choose $mL>0$. Normalizability 
of the solution requires that
\begin{eqnarray}
mL\geq\frac 1 2: \qquad &\psi&\propto z^{\frac 3 2 +mL}\xi \quad\implies\quad \xi_{\alpha}=-i \sigma^3_{\alpha\dot{\alpha}}\bar{\xi}^{\dot{\alpha}}\nonumber\\
0\leq mL< \frac 1 2: \qquad &\psi&\propto z^{\frac 3 2 \pm mL}\xi \quad\implies\quad \xi_{\alpha}=\mp i \sigma^3_{\alpha\dot{\alpha}}\bar{\xi}^{\dot{\alpha}}
\label{adsfermion}
\end{eqnarray}
and again chiral symmetry is necessarily broken. Note that for the AdS case,
there is no freedom to chose the phase $\varphi$. This is
basically because the bulk mass operator itself plays the r\^{o}le of
a boundary mass term. This is easily seen by performing a Weyl
rescaling, $\psi =(z/L)^{3/2}\chi$: the lagrangian for $\chi$ is
just given by eq.~(\ref{flat}), but with a position dependent mass
$m\to ML/z$, which blows up at $z=0$. The exponent in the
asymptotic behavior is precisely the index $E$ of the corresponding 
representation. Note that for $mL<1/2$, two inequivalent 
boundary conditions are possible, corresponding to a double quantization, as happens
for scalars in AdS \cite{breit}. The existence of one and two solutions respectively 
for $mL\geq1/2$ and $0\leq mL <1/2$, nicely matches eq.~(\ref{fermion}) and the unitarity bound $E\geq 1$.

In the QED case, the boundary condition (\ref{adsfermion}) for a single charged spinor 
would break electric charge; in order to conserve electric charge, the boundary conditions
must relate $\psi_+$ to $\bar \psi_-$.\footnote{Indeed, for $mL>1/2$ 
the charge preserving boundary condition is forced on us by normalizability. 
For $0\leq mL<1/2$, compatibly with normalizability, there exist two other, inequivalent, 
charge-breaking boundary conditions. We will consider these other possibilities elsewhere.} 
Repeating the exercise above with the two spinors, normalizability of the solutions requires
\begin{eqnarray}
mL\geq\frac 1 2: \qquad &\psi_-,\psi_+&\propto z^{\frac 3 2 +mL} \quad\implies \quad \psi_{+\alpha}=-i \sigma^3_{\alpha\dot{\alpha}}\bar{\psi}_-^{\dot{\alpha}}\label{fermionads}\nonumber \\
0\leq mL< \frac 1 2: \qquad &\psi_-,\psi_+&\propto z^{\frac 3 2 \pm mL}\quad\implies\quad \psi_{+\alpha}=\mp i \sigma^3_{\alpha\dot{\alpha}}\bar{\psi}_-^{\dot{\alpha}}
\end{eqnarray}

Given the boundary conditions for the fermions, supersymmetry then determines 
the boundary conditions for the scalars.  By acting with the unbroken 
supersymmetries (\ref{unbrokensusy}) on the fermionic boundary conditions, one finds
\begin{eqnarray}
mL\geq\frac 1 2: \qquad && z\to 0  \quad\implies\quad\phi_+= \phi_-^*\left[1+O(z)\right ]\nonumber\\
0\leq mL< \frac 1 2: \qquad && z\to 0  \quad\implies\quad\phi_+=\pm  \phi_-^*\left[1+O(z)\right ]
\label{scalarads}
\end{eqnarray}
where the sign in the second eq. is correlated with the sign for the fermions.
We can see that this is consistent  with the equations of motion for the scalars: 
In the scalar sector, by solving the wave equation for the two
mass eigenstates, $\phi_1$ and $\phi_2$, we find that
\begin{eqnarray}
\lim_{z\to 0}\phi_1&=&z^{2+mL}A_2(x)+z^{1-mL}B_2(x)\nonumber \\
\lim_{z\to 0}\phi_2&=&z^{1+mL}A_1(x)+z^{2-mL}B_1(x)\,.
\end{eqnarray}
For $mL>1/2$, normalizability alone implies that $B_1=B_2=0$, corresponding to the first 
solution in eq. (\ref{scalarads}). For $m L<1/2$, the mass of the two scalars is in the range 
where double quantization is allowed, and so we can choose $A_1=A_2=0$ (consistently with
supersymmetry), corresponding to the second solution in (\ref{scalarads}). Note that, as a combined effect of the boundary 
conditions for fermions and scalars, $R-$symmetry is broken in the matter sector even for $m=0$.

Finally, we can also fix the boundary condition for the vector multiplet.
By taking Neumann boundary conditions for the gauge field and acting with the supersymmetry 
transformations, we find that the appropriate sign of the gaugino boundary condition is
\begin{equation}
\partial_z F^{\mu\nu}\big \vert_{z=0}=0\qquad F^{\mu 3}\big \vert_{z=0}=0\qquad \lambda_\alpha=i\sigma^3_{\alpha\dot\alpha}\bar\lambda^{\dot\alpha}\big \vert_{z=0}\label{gaugeboundary}
\end{equation}

To summarize, the presence of the boundary in AdS$_4$ always breaks chirality 
and $R-$sym\-me\-try, even when there is no source of explicit breaking in the bulk action.
The physics is essentially that of half of flat space. What is special to
AdS$_4$ is that the chiral symmetry is broken, while the maximal number 
of isometries is preserved. This is, of course, crucial to give a  meaning 
to a mass smaller than the curvature of the space.

\section{Gaugino Mass}
\label{mass}

The boundary conditions derived above provide the necessary `mass insertions' to
give rise to an IR contribution to the gaugino mass. Focussing on the case of massless SQED,
let us now compute the gaugino mass at 1-loop order. 

\subsection{Chiral breaking correction to the self energy}

The computation is particularly transparent in the case of massless matter,
where the chiral symmetry breaking is entirely due to the boundary effects. 
(We present the massive case in the appendix.) When $m=0$, the chiral matter 
supermultiplet is conformally coupled. As a consequence, the full SQED 
action in this case is invariant under Weyl transformations at the classical 
level. This allows us to map the theory in AdS space to one living on 
half of flat space and perform all the computations using familiar flat space formulae. 
This is achieved via the superconformal rescaling
\begin{eqnarray}
\phi&=&\left(\frac z L\right) \hat{\phi},~~~~~
\psi=\left(\frac z L\right)^{\frac 3 2} \hat{\psi}, ~~~~~
\lambda=\left(\frac z L\right)^{\frac 3 2} \hat{\lambda}, ~~~~~
A_M = \hat{A}_M,\label{rescaling0}\\
s&=&\left(\frac z L\right) \hat s, ~~~~~u=\left(\frac z L\right)^2 \hat u,~~~~~g_{MN}=\left(\frac z L\right)^2 \hat g_{MN}.
\label{rescaling}
\end{eqnarray}
After the rescaling, $\hat g_{MN}\equiv \eta_{MN}$ and $\hat{S}=(L/z)(1+\theta^2/z)$. 
Since SQED is Weyl invariant (at tree level), the compensator decouples, and 
we are left with the tree level action for massless, SQED in half of flat space, with a boundary at
$z=0$. The boundary conditions on the fields are most easily
implemented by performing an orbifold projection. From the results 
in the previous section, we have (dropping the circumflexes on the fields ),
\begin{eqnarray}
\psi_+(X)&=&-  i\sigma^3 \bar{\psi}_-(\tilde{X}),\nonumber \\
\phi_+(X) & =&  \phi_-^*(\tilde{X}), \nonumber \\
A_\mu(X) &=&  A_\mu(\tilde{X}), \nonumber\\
A_z(X)& = &-  A_z(\tilde{X}), \nonumber\\
\lambda (X) & = & i\sigma^3
\bar{\lambda}(\tilde{X}),
\label{orbifold}
\end{eqnarray}
where $\tilde{X}=(x, -z)$ is the position of the image point.
The flat space propagators can be written down directly using the method of image
charges. For the scalars, one has
\begin{eqnarray}
\langle \phi_+(X_1) \phi_+^*(X_2)\rangle& = &\langle \phi_-(X_1) \phi_-^*(X_2)\rangle= \frac 1 {4 \pi^2} \frac 1
{(X_1-X_2)^2+i\epsilon}, \\
\langle \phi_+(X_1) \phi_-(X_2)\rangle& = &\langle
\phi_-^*(X_1) \phi_+^*(X_2)\rangle= \frac 1 {4
\pi^2} \frac 1 {(X_1-\tilde{X}_2)^2+i \epsilon}. \nonumber
\label{scalarprop}
\end{eqnarray}
Similarly, for the fermions, 
\begin{eqnarray}
\langle
\psi_{+\alpha}(X_1)\bar{\psi}_{+\dot\beta}(X_2)\rangle&
= &\langle
{\psi}_{-\alpha}(X_1)\bar{\psi}_{-\dot\beta}(X_2)\rangle=
\frac i {2 \pi^2} \frac{(X_1-X_2)_M
\sigma^M_{\alpha\dot{\beta}}}
{[(X_1-X_2)^2+i\epsilon ]^2}, \\
\langle
\psi_{+\alpha}(X_1){\psi}_{-}^{\beta}(X_2)\rangle& =
&-\frac 1{2 \pi^2} \frac{(X_1-\tilde{X}_2)_M (\sigma^M\bar{\sigma}^3)_\alpha^\beta}
{[(X_1-\tilde{X}_2)^2+i\epsilon ]^2}. \nonumber
\label{fermionprop}
\end{eqnarray}
One can see that the $i\epsilon$ prescription in Feynman's propagator selects 
implicitly boundary conditions at $z=\infty$: these are the Hartle-Hawking 
boundary conditions, appropriate to the Poincar\'e patch \cite{giddings}.

\begin{figure}[t!]
\begin{center}
\includegraphics[width=7cm]{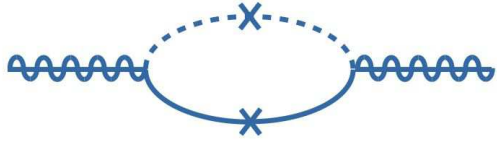}
\caption{Chiral breaking  1-loop correction to the gaugino self energy.
The ``mass'' insertions correspond to boundary effects. }
\label{diag}
\end{center}
\end{figure}

The off-diagonal propagators determine the chiral-breaking
contribution to the gaugino self-energy  in Fig. \ref{diag} 
\begin{eqnarray}
\Sigma_{\alpha}^{\ \beta}(X_1,X_2) &=& i\langle J_{\alpha}(X_1)J^{\beta}(X_2) \rangle,
\label{chiralself}
\end{eqnarray}
where $J_{\alpha} = i \sqrt{2} g (\phi_+^* \psi_{+\alpha} -
\phi_-^* \psi_{-\alpha})$ and where our convention on the
self-energy is defined by  $\Gamma_{1PI}\supset
\int \frac{1}{2}\lambda^\alpha(X_1)\Sigma_\alpha^{\
\beta}(X_1,X_2)\lambda_\beta(X_2)$. Performing  the Wick contractions, we have
\begin{eqnarray}
\Sigma_{\alpha}^{\ \beta}(X_1,X_2)&=& 4ig^2 \langle \phi_+^*(X_1) \phi_-^*(X_2) \rangle \langle \psi_{+\alpha}(X_1)\psi_{-}^{\beta}(X_2) \rangle, \nonumber \\
&=&-\frac{ig^2}{2 \pi^4 } \frac{(X_1-\tilde{X}_2)_M (\sigma^M\bar{\sigma}^3)_\alpha^\beta}
{[(X_1-\tilde{X}_2)^2+i\epsilon ]^3}.
\label{selfenergy}
\end{eqnarray}
Notice that this contribution is non-local, and comes from long-distance physics, 
as opposed to eq.~(\ref{gauginobilinear}). In order to extract from 
$\Sigma_{\alpha}^{\ \beta}(X_1,X_2)$ the correction to the gaugino mass, we must evaluate it 
on a solution of the massless (tree level) wave equation. This is the analogue
of computing the self-energy at zero momentum in flat space.
The general solution of the bulk Dirac equation for a massless gaugino is
\begin{equation}
\lambda_0(X)= e^{ip_MX^M}\xi, \qquad \bar\sigma^M p_M \xi=0, \qquad p^Mp_M=0\,.
\label{dirac4}
\end{equation}
Physical states must also satisfy the boundary condition in eq.~(\ref {gaugeboundary}). In order to achieve that, 
two solutions with opposite velocity, $p^3/p^0$, in the $z$-direction should be superimposed. 
However, as we shall explain in a moment, the correct procedure we must follow in 
the Poincar\'e patch in order to study the 1-loop corrected wave equation is to work with solutions 
of the Dirac equation that satisfy boundary conditions at the horizon $z\to \infty$ rather than at the boundary 
$z=0$. This is  closely related to the AdS/CFT prescription. Alternatively we could overcome this issue 
by performing an euclidean computation, as in this case Poincar\'e co-ordinates cover the whole space, 
but we find it more physical to address directly the Lorentzian point of view.

To obtain the IR contribution to the gaugino mass, we must convolute eq.~(\ref{selfenergy}) 
with (\ref{dirac4}). We thus find,
\begin{eqnarray}
\int d^4X_2\, \Sigma_\alpha^{\ \beta}(X_1,X_2)\lambda_{0\beta}(X_2)&=&-\frac {ig^2}{ 8\pi^4}\int \, d^4X_2  \frac {\partial}{\partial \tilde{X}_2^M}\left(\frac{(\sigma^M\bar{\sigma}^3)_\alpha^\beta}
{[(X_1-\tilde{X}_2)^2+i\epsilon ]^2}\right)\lambda_{0 \beta}(X_2)\nonumber\\
&=&\frac {i g^2}{8 \pi^4}\int d^3 x_2
\frac{1}{[(x_1-x_2)^2+z_1^2+i\epsilon ]^2}e^{ip_\mu x_2^\mu} \xi_{\alpha}
\label{integral}
\end{eqnarray}
where in the last step we integrated by parts and used $ \bar\sigma^M \partial_M \lambda_0=0$. 
In the resulting boundary integral, we used the explicit expression for $\lambda_0$ in (\ref{dirac4}). 
Notice that $x$ are coordinates on the boundary.  Performing the last integral explicitly we thus find,
\begin{equation}
\frac{1}{2}\int \, d^4X_2\, \Sigma_\alpha^{\ \beta}(X_1,X_2)\lambda_{0\beta}(X_2)=\frac {g^2}{16\pi^2} \frac 1 {z_1}e^{i(p_\mu x_1^\mu + |p| z_1)  }\xi_{\alpha},
\label{finite}
\end{equation}
where the $i\epsilon$ in the original integral fixes the sign of $p_3=\sqrt{-p_\mu p^\mu}$ to be positive.
The right hand side of eq. (\ref{finite}) is proportional to the original spinor if this satisfies the
Hartle-Hawking boundary conditions: positive frequencies purely outgoing and negative frequencies purely incoming. 
This means that when evaluated on this class of solutions of the bulk Dirac equation, the IR contribution 
to the self-energy $\Sigma_{\alpha}^{\ \beta}$, acts like a mass term $m_{IR}$ which is precisely equal and 
opposite to the anomaly mediated contribution (see eq.~(\ref{gauginobilinear}) after performing the Weyl rescaling in eqs.~(\ref{rescaling0}),(\ref{rescaling})). Thus an exact cancellation between UV and IR effects arises, 
as promised in eq.~(\ref{uvir}).  It is the clever relation among these two contributions 
that ensures the masslessness of the gaugino, as demanded by supersymmetry. 
This is the main result of our paper.

It remains to be explained why our computation works only for the class of solutions of the form (\ref{finite}). 
These solutions correspond to the creation of incoming particles at the past horizon $H^-$
and to the destruction of outgoing particles at the future horizon $H^+$ that separate the Poincar\'e patch from the rest of AdS. 
Intuitively such processes can be described by causality using solely the fields in the Poincar\'e patch. 
Other solutions correspond to processes that are not captured by the Poincar\'e patch alone and probe other regions of
global AdS. In this case there will be extra-contributions from the rest of the space and a computation 
in global coordinates would be required. That such contributions exist follows from 
the fact that the Feynman propagator is non-vanishing between a point inside the Poincar\'e 
patch and one outside. Had we worked in global coordinates we could have directly checked that the 
cancellation of the gaugino mass occurs for arbitrary physical states (i.e. solutions of the wave equation
that satisfy the boundary conditions). 

Our result can however be readily interpreted from the viewpoint  of the AdS/CFT correspondence \cite{adscft}.
Even though Lorentzian AdS/CFT is not nearly as  developed as on Euclidean space, we do not see obvious obstructions in the case at hand.\footnote{Indeed it is to be expected that, just as there is a procedure to analytically continue 
a CFT from Euclidean to Lorentzian space, there should also exist a similar procedure 
to  analytically continue the correspondence from Euclidean to Lorentzian AdS. 
At least in some simple cases this was outlined for instance in Refs. \cite{giddings,Skenderis:2008dh}.} 
 From this perspective, the boundary field combination 
\begin{equation}
\lambda_-(x)=\lambda(x)-i\sigma^3\bar\lambda(x)
\end{equation}
should be viewed as an external source probing the system (the dual CFT).
 Notice that $\lambda_-$ is precisely the combination 
that is set equal to zero for the AdS quantum fields. 
Performing a path integral over the bulk fields with vacuum boundary conditions at $H^{\pm}$ one 
obtains a functional $Z(\lambda_-)$ which generates the correlators of the associated dual operator 
in the CFT. Given $\lambda_-$, a classical source localized at the boundary, the choice of initial 
and final vacuum states for our path integral  fixes the boundary condition for the corresponding 
bulk field at $z\to \infty$. Working with  plane waves, this prescription corresponds precisely to the 
Hartle-Hawking boundary condition we encountered previously. This gives a prescription for finding a unique extension 
of $\lambda_-$ into the bulk, by requiring that its (effective) action be stationary. 

At tree level, we have the boundary effective action
\begin{equation}
\ln Z= S_{bd}=-\frac{1}{4}\int d^3x \left (\lambda\lambda+\bar\lambda\bar\lambda\right )=\int d^3x \lambda_-\frac {\sigma^3\bar \sigma^\mu }{\sqrt {\partial^2+i \epsilon}}\partial_\mu\lambda_-,
\end{equation}
corresponding to the correlator of a  dual fermionic current of scaling dimension $\frac{3}{2}$:
\begin{equation}
\langle O_\alpha(x) O^\beta (0)\rangle = \frac{ x^\mu(\sigma^3\bar\sigma_\mu)_\alpha^\beta}{(x^2+i\epsilon)^2}.
\end{equation}
The 1-loop computation we have performed is directly translated into a 1-loop computation 
of the boundary effective action. The only difference from before is that we need to consider also
solutions with Euclidean boundary momenta $p_\mu p^\mu>0$. In this case the solution 
in the bulk corresponds to the unique regular solution at $z\to\infty$ as prescribed by Euclidean AdS/CFT.
Needless to say the previous computation can be continued to the Euclidean region so that the self energy 
is diagonal on these solutions. Working at 1-loop accuracy, the corrected boundary effective action is simply 
obtained by substituting the tree level bulk solution into the 1PI bulk effective action. However 
our previous result was precisely that the total (UV + IR) 1PI vanishes on the very solution of the 
massless Dirac equation that satisfied  the AdS/CFT boundary conditions at $z\to \infty$ (that is with 
the same exponent as in eq.~(\ref{finite})). Thus we conclude that at the 1-loop level the boundary 
action is unaffected and thus the dimension of the CFT operator dual to the gaugino field is not 
renormalized, consistently with supersymmetry. 

What we have learned is an amusing lesson on the r\^ole of the anomaly mediated gaugino mass. 
The basic reason for its existence is that AdS$_4$ behaves as  2+1-dimensional field theory as far as chirality is
concerned. The mass of fermions is thus additively renormalized by calculable boundary effects. 
On the other hand, supersymmetry mandates the gaugino to be exactly massless. The simple SQED case, 
in the end, shows that the only way to achieve this is via the existence of suitable short distance effects, 
in one-to-one correspondence with the long distance effects. This is yet another illustration of the UV 
insensitivity of anomaly mediation.

\subsection{Chiral preserving correction: wave function renormalization}

In the previous section we have shown that the chiral breaking part in the 1-loop self energy 
does not correct the gaugino mass, nor, similarly, does it correct the boundary effective action. 
However, strictly speaking there is yet another contribution to the gaugino self-energy
that we need to consider. This is   the `chirality-preserving' contribution, $\Sigma_{\alpha \dot{\beta}}$,  
the one associated with wave-function renormalization. 
The issue at hand arises even in the absence of supersymmetry. 
We will show that this contribution vanishes when acting on a 
massless spinor. This result may seem obvious at first sight, 
based on our usual flat space intuition. Indeed, in flat Minkowsky space, Lorentz
invariance constrains this term to be proportional to $f(\Box)\,
\!\!\not \!\! \,
\partial$, which vanishes on-shell as long as $f$ is not too
singular (in fact, $f$ is a logarithmic function). However, the
situation is more subtle in AdS, since, at the quantum level, the
boundary makes itself felt even inside the bulk, and therefore the
$z$ direction is not manifestly equivalent to the others. The
purpose of this section is to clarify this issue. An extra
complication comes from the need to regularize the divergent part
of $\Sigma_{\alpha \dot{\beta}}$. We shall again focus on massless
SQED, for which we can work in the conformally rescaled basis 
(\ref{rescaling}). The general case is briefly considered in the appendix.  
Working in position space, we find it convenient to use the method of differential
regularization \cite{Freedman:1991tk}.

The unregulated $\Sigma_{\alpha \dot{\beta}}$ is given by,
\begin{eqnarray}
\Sigma_{\alpha \dot{\beta}}(X_1,X_2)&=& i\langle J_{\alpha}(X_1)J_{\dot{\beta}}(X_2) \rangle\nonumber \\
&=&-4i g^2 \langle \phi_+(X_1) \phi_+^*(X_2) \rangle \langle \psi_{+\alpha}(X_1)\bar{\psi}_{+\dot{\beta}}(X_2) \rangle
\end{eqnarray}
This corresponds to the following correction to the effective action
\begin{equation}
\Gamma =-\frac{g^2}{2\pi^4} \int d^4X_1d^4X_2 \bar \lambda
(X_1)
\frac{{X_{12M}} \bar \sigma^M}{(X_{12}^2+i\epsilon)^3}\lambda(X_2),
\end{equation}
where $X_{12}=(X_1-X_2)_M$.  This expression has, however, a non-integrable singularity at $X_{12}=0$, 
which must be regulated. Na\"{\i}vely, using differential regularization amounts
to replacing
\begin{equation}
\frac{{X_{12M}}}{(X_{12}^2+i\epsilon)^3}\to
\frac{1}{16} \frac{1}{\partial X_1^M}\left(\Box_1\frac{\ln (X_{12}^2M^2)}{X_{12}^2+i\epsilon}\right )\, ,
\label{reg}
\end{equation}
where $M$ plays the r\^{o}le of the renormalization mass scale.
This cannot, however, be the full story, since the explicit mass
scale $M$ breaks  dilatation invariance $X \rightarrow kX$. In the
rescaled basis, $SO(3,2)$ arises as the  subgroup of $SO(4,2)$ which
is left unbroken by the compensator background $\tilde s=L/z$
\cite{fubini}.  Consequently the regulated self-energy in
eq.~(\ref{reg}) does not respect the AdS isometries. As the lack
of invariance follows from the regularization, the counterterm
needed to restore the symmetry must be local, and must of course
involve the compensator. By simple reasoning one can quickly
derive the unique form of this counterterm. In order to do so, let us
imagine that we had regulated the loop in a manifestly covariant
fashion, by introducing Pauli Villars fields with mass $M$. The
crucial aspect of Pauli-Villars fields is that, being massive,
their quadratic lagrangian depends directly on the compensator,
$\tilde s$, via  the substitution
\begin{equation}
M\to M\times \tilde s (z)=M\times \frac{L}{z},
\label{modify}
\end{equation}
which formally restores conformal invariance. However it does not
make any sense to simply perform this replacement in eq.~(\ref{reg}).
To find out how  eq.~(\ref{reg}) is modified we must be a tad more
careful. We just need to focus on  the $M$-dependent part of the
regulated self-energy. Using the identity
\begin{equation}
\Box \frac{1}{x^2+i\epsilon}=4\pi^2 i\delta^4(x),
\end{equation}
the $M$-dependent part of the effective action is given by
\begin{equation}
\Delta \Gamma_{UV}=-\frac{ig^2}{8\pi^2} \ln M^2\int d^4 X\bar
\lambda(X)\bar \sigma^M\partial_M \lambda(X), \label{wfuv}
\end{equation}
whose unique local covariantization is\footnote{Indeed, compatibly
with locality and power counting,  another term is na\"{\i}vely
possible:
\begin{equation}
[\partial_z\ln \tilde s (z)]\bar \lambda\bar\sigma^3\lambda\,.
\end{equation}
 This term must however  be discarded as it explicitly breaks CP (the regulated theory is formally CP-invariant, even though parity is, of course, `spontaneously' broken by the expectation value of $\tilde s$).}
\begin{equation}
\Delta \Gamma_{UV}=-\frac{ig^2}{8\pi^2}  \ln( M \tilde s (z) )\left [\bar \lambda\bar \sigma^M\partial_M \lambda - \partial_M \bar \lambda\bar \sigma^M \lambda\right ].
\label{amwf}
\end{equation}

The local $\ln \tilde s$ term gives the following correction to
$\delta \Gamma/\delta \bar\lambda$
\begin{equation}
-\frac{ig^2}{8\pi^2} \partial_M \ln \tilde s
\bar\sigma^M\lambda=\frac{ig^2}{8\pi^2} \frac{1}{z} \bar \sigma^3\lambda.
\label{uv}
\end{equation}
On the other hand, from eq.~(\ref{reg}) the `IR' contribution to the equation of motion is
 \begin{equation}
\frac{g^2}{2\pi^4} \frac{1}{16} \int d^4X_2 \  \frac{\partial}{\partial X_2^M}\left(\Box_1\frac{\ln (X_{12}^2M^2)}{X_{12}^2}\right ) \bar \sigma^M \lambda(X_2).
\label{nonlocwave}
\end{equation}
To investigate how this non-local contribution affects the gaugino mass we must compute it 
on the solution $\lambda_0$ of the massless wave equation
specified by (\ref{dirac4}). Integrating by parts and using
$\bar \sigma^M\partial_M \lambda_0=0$, eq.~(\ref{nonlocwave}) becomes
\begin{align}
-\frac{g^2}{32\pi^4} \int d^3x_2 \left (\Box_1\frac{\ln M^2 X_{12}^2}{X_{12}^2}\right ) \bar \sigma^3 \lambda_0(X_2)\big |_{z_2=0} &=
\frac{g^2}{8\pi^4}\int d^3x_2\frac{1}{(X_{12}^2+i\epsilon)^2} \bar\sigma^3 \lambda_0(X_2) \nonumber\\
 =-\frac{ig^2}{8\pi^2} \frac{1}{z_1} \bar \sigma^3\lambda_0(X_1), \label{ir}
\end{align}
where the final integral is identical to the one computed in the previous section,
eq. (\ref{integral}). Again the last identity is only valid for solutions satisfying the Hartle-Hawking boundary conditions. 
We thus find that the contributions in eqs. (\ref{uv}) and (\ref{ir}) again cancel so that the propagation of the gaugino 
is not affected. In particular the gaugino remains massless. Note that, while the cancellation in the
previous section relies on supersymmetry, this effect is independent of supersymmetry. This cancellation between UV and IR contributions, dictated by the AdS isometry (a subgroup of the conformal group), can be viewed as an $N=0$ counterpart of the one found previously. This is perhaps not surprising, as anomaly mediation is itself the 
supersymmetric counterpart of the trace anomaly. Indeed, in a superfield formalism, these two separate cancellations 
would be manifestly related.

\section{Summary}
\label{out}

We  studied the r\^ole played by anomaly mediated (AM) mass terms
in $N=1$ theories on AdS$_4$  with unbroken supersymmetry. 
For simplicity we focussed on the gaugino mass term in SQED with massless matter. 
We  showed that the AM gaugino mass term is required by the 
super-AdS algebra in order to exactly cancel another 1-loop contribution, of infrared 
origin and associated with the AdS boundary. The latter effect originates because 
chirality ($R-$symmetry in this case) is necessarily broken by reflection at a 2+1-dimensional boundary. 

Indeed, by computing first this finite IR effect (which does not require the introduction of a regulator) 
and by using the fact that the algebra dictates a massless gaugino, we could have argued the need 
for a local, UV generated, AM contribution. Since the latter is independent of whether 
the theory lives in flat or curved space, that would have provided yet another derivation of 
AM gaugino masses. The possibility of relating the AM mass to purely 
IR quantities illustrates the ``UV insensitivity'' of this effect, a property which makes it potentially 
relevant in phenomenological applications. The fact that AM effects represent local parts of non-local 
structures in the 1PI action is well known. Our result provides a new twist on that perspective: 
the AM gaugino mass is just a reflection of the breakdown of chirality
at the 2+1-d boundary of AdS$_4$.

There are several directions in which one might extend and improve our result. One 
obvious possibility is to perform the same computation in the non-abelian case, where, 
unlike in the abelian case, proper gauge-fixing will be needed. Another problem concerns 
the r\^{o}le of all other AM terms, such as sfermion masses and ``A-terms":  
it should be possible to derive them from consistency conditions as well, 
but probably in a more subtle way than for the gaugino mass.

In this paper we worked on the Poincar\'e patch. This procedure is clean for the euclidean 
case and from the AdS/CFT standpoint:  our computation corresponds to checking that, 
as expected by supersymmetry, the scaling dimension  of  the operator dual to the gaugino 
field is not renormalized. The Lorentzian computation is more delicate, as we have to deal 
with boundary conditions at the horizons which separate the chosen patch from the rest of AdS. 
It would then be interesting to try to perform the same computation in global coordinates, 
and check that, in that case, the 1-loop self energy does vanish when convoluted with 
the normalizable solutions. Finally, it would be interesting to understand the r\^ole of anomaly 
mediation purely from the CFT viewpoint. The AdS bulk picture is that the gaugino must 
be massless even though chirality is broken, corresponding to non-vanishing 
$\Sigma_{\alpha}^{\ \beta}$ off-shell. In the CFT picture, the non-vanishing of $\Sigma_{\alpha}^{\ \beta}$, 
shows up in the 4-point function of operators dual to the AdS matter fields. However
it is not immediately obvious how to translate the bulk picture to the boundary, 
since there is no notion of chirality in 2+1-d field theory.

\vspace{1cm} {\bf Acknowledgments} We would like to thank M.
Bianchi, S. Giddings, M. Porrati, S. Sybiriakov, A. Wulzer and A. Zaffaroni for
useful discussions. HK was supported by the CQUeST of Sogang University
with grant number R11-2005-021 and CS by the Swiss National 
Science foundation. The work of R.R. is partially supported by the Swiss National Science  
Foundation under contract No. 200021-116372. R.R. thanks the Aspen Center for Physics 
where part of this work was carried out.

\appendix

\section*{Appendix: Massive Charged Matter}
\renewcommand{\theequation}{\thesection.\arabic{equation}}
\setcounter{section}{1}
\setcounter{equation}{0}

The cancellation of the UV and IR contributions to the gaugino
mass, being a consequence of the algebra, is a general effect
which must hold for any mass of the matter fields. In this
appendix we check explicitly the cancellation for arbitrary values
of $m$ in the superpotential. This computation can also be
interpreted as the derivation of the anomaly mediated UV
contributions (\ref{gauginobilinear}), (\ref{amwf}) using Pauli-Villars
fields.

For massless matter, the only source of chiral symmetry breaking
is due to the presence of the boundary, while when $m\ne0$, chiral
symmetry is broken also in the bulk. In this case, the matter is
not conformally-coupled and, therefore, the propagators cannot be
obtained by simply rescaling the flat space results. A full AdS
computation is required. 

We will need the propagators for a chiral multiplet with arbitrary mass.
The scalar propagator associated to the representation $D(E,0)$
($(mL)^2=E(E-3)$) is given by\footnote{This formulae hold for
$E>3/2$ where both scalars in the chiral multiplet have standard
boundary conditions. This is the range where a single quantization is possible.}
\begin{eqnarray}
\Delta(E,0) &=& \frac {1} {(4\pi)^2 L^2} \frac
{\Gamma[E]\Gamma[E-1]}{\Gamma[2E-2]} \left(\frac 2 u\right)^{E}\,
_2F_1\left(E,E-1;2E-2 , -\frac 2 {u}\right), \nonumber
\end{eqnarray}
where we have introduced the AdS invariant length,
\begin{equation}
u= \frac {(X_1-X_2)^2+i \epsilon}{2 z_1 z_2},
\end{equation}
The fermion propagator associated to the representation $D(E+1/2,1/2)$
can be found in Ref. \cite{allen},
\begin{eqnarray}
\langle \psi_{+\alpha}(X_1)\psi_{-}^{\beta}(X_2) \rangle & =&
\frac
{-\Gamma[E]\Gamma[E+1]}{(32 \pi^2 L^3)\Gamma[2E-1]}\left(\frac
{2}{u+2}\right)^{E+1} \, _2F_1\left(E+1,E-1; 2 E-1,\frac 2
{u+2}\right)\times \Gamma_{\alpha}^{\ \beta}, \nonumber \\
\langle \psi_{\pm\alpha}(X_1)\bar{\psi}_{\pm\dot{\beta}}(X_2)
\rangle & =& \frac
{i \Gamma[E]\Gamma[E+1]}{(32\pi^2 L^3)\Gamma[2E-1]}\left(\frac
{2}{u+2}\right)^{E+1} \, _2F_1\left(E+1,E; 2 E-1,\frac 2
{u+2}\right)\times \Gamma_{\alpha\dot{\beta}}\nonumber \\
\end{eqnarray}
where,
\begin{eqnarray}
\Gamma_{\alpha}^{\ \beta}&=&\frac {(X_1-\tilde{X}_2)_M
(\sigma^{M}\bar{\sigma}^3)_{\alpha}^{\
\beta}}{\sqrt{z_1 z_2}}
\nonumber \\ \Gamma_{\alpha\dot{\beta}}&=&\frac {(X_1-X_2)_M
\sigma^M_{\alpha \dot{\beta}}}{\sqrt{z_1 z_2}}
\end{eqnarray}

As in the massless case, the contribution of the matter loop to
the gaugino mass arises from the the self-energy
(\ref{chiralself}),
\begin{equation}
\Sigma_{\alpha}^{\ \beta}(X_1,X_2)= 4 i g^2\langle \phi_+^*(X_1)
\phi_-^*(X_2) \rangle \langle
\psi_{+\alpha}(X_1)\psi_{-}^{\beta}(X_2) \rangle
\end{equation}
where now
\begin{eqnarray}
\langle \phi_+(X_1)\phi_-(X_2)\rangle & =& \frac
{\Delta(E+1,0)-\Delta(E,0)} 2,
\end{eqnarray}
and the fermion belongs to the representation $D(E+1/2,1/2)$.

In order to compute the contribution to the gaugino mass, we evaluate the self-energy 
on the solution of the massless gaugino equation as in section 3.1. This highly non-trivial integral of
hypergeometric functions can be evaluated numerically by choosing
the simplest solution of the massless equation of motion,
$\lambda_0(X_1)=z^{3/2}\xi_0$,
\begin{equation}
\int dX_2 \sqrt{-g}  \Sigma_{\alpha}^{\ \beta}(X_1,X_2) \lambda_{0
\beta}(X_2)=-\frac {g^2}{8\pi^2 L} \lambda_{0\alpha}(X_1).
\end{equation}
Following the discussion in section 3.1 we expect the same to hold 
for any solution satisfying the appropriate boundary conditions.
This contribution as expected does not depend on the mass and cancels
the anomaly-mediated UV contribution, proving for general $m$ that
this term is necessary for the consistency of the supersymmetric
theory. As a check of this result, one can consider the limit $m
\gg 1/L$, as done in \cite{ds}. In this limit, the curvature is a
small effect and the loop can be computed using flat-space
propagators, but with the AdS mass splitting.

For completeness we also checked the wave functions contribution.
The chiral preserving contribution to self-energy in general reads,
\begin{equation} \Sigma_{\alpha\dot\beta}(X_1,X_2)=-
2ig^2\left[\langle \phi_1(X_1) \phi_1^*(X_2) \rangle+\langle
\phi_2(X_1) \phi_2^*(X_2) \rangle\right] \langle
\psi_{+\alpha}(X_1)\psi_{+\dot{\beta}}(X_2) \rangle
\end{equation}
Repeating the same steps as in section 3.2, we find numerically,
\begin{equation}
\int  dX_2  \sqrt{-g} \Sigma_{\alpha\dot{\beta}}(X_1,X_2)
\lambda^{\dot{\beta}}(X_2)=-\frac {g^2}{8\pi^2 L}
\lambda_{\alpha}(X_1).
\end{equation}
independently of the mass. This calculation also proves that by
regulating the theory with Pauli-Villars fields there is an $N=0$
anomaly mediated contribution of the form considered before. In
this case the contribution of the heavy fields with $m \gg 1/L$
cannot be obtained with the flat space propagators since this
effect is entirely due to the fact that the theory lives in curved space.


\end{document}